\documentclass{svproc}

\usepackage{url,color}

\usepackage{amsmath,amssymb,mathrsfs}
\usepackage{enumerate}
\usepackage{graphicx,subfigure,hhline}

\definecolor{ggreen}{cmyk}{1,0,1,0}

\begin{document}
	\mainmatter             
	
	\title{Investigating dynamics and asymptotic trend to equilibrium 
		in a reactive BGK model }
	
	
	
	\titlerunning{Asymptotic trend in a reactive BGK model}  
	
	\author{G. Martal\`o\inst{1} \and A.~J. Soares\inst{2} \and R. Travaglini\inst{3,1}}
	
	\authorrunning{G. Martal\`o et al.} 
	
	\tocauthor{Giorgio Martal\`o, Ana Jacinta Soares, Romina Travaglini}
	\institute{Department of Mathematical, Physical and Computer Sciences\\ 
		University of Parma, Parco Area delle Scienze 53/A,  43124,
		Parma, Italy \\
		\email{giorgio.martalo@unipr.it}
		\\
		\and
		Centre of Mathematics, University of Minho\\
		Campus de Gualtar,
		4710-057 Braga, Portugal\\
		\email{ajsoares@math.uminho.pt}
		\\              
		\and
		Istituto Nazionale di Alta Matematica ``Francesco Severi''\\
		Piazzale Aldo Moro 5, 00185, Roma, Italy\\
		\email{romina.travaglini@unipr.it} 
	}
	
	\maketitle              
	
	\begin{abstract}
		We investigate numerically a recent BGK-type  model for a multi-component mixture of monatomic gases, 
		undergoing a reversible bimolecular chemical reaction. The model replaces each collisional term of the Boltzmann equation with a relaxation term, thereby describing separately the effects of the mechanical processes and the chemical reaction. Additionally, the model exhibits consistency properties. The correct entropy production is ensured when auxiliary temperatures in the chemical contributions share a common value. We assume isotropic distributions and perform numerical simulations for the macroscopic fields to appraise how the dynamics push the mixture toward thermalization and chemical equilibrium. We show that the hypothesis on the equalization of fictitious species temperatures is justifiable to ensure the monotonicity of the classical $H$-Boltzmann functional. Simulations show that, when initial temperatures are far from equilibrium, the relaxation towards equilibrium occurs at a later stage and the classical $H$-Boltzmann functional is not monotone during the initial transient.
		\hspace*{4cm}
	\end{abstract}
	
	
	\section{Introduction}
	\label{sec:int}
	
	Kinetic theory and its descriptions based on Boltzmann equations provide a natural framework to describe the evolution of rarefied gas mixtures. Unfortunately, the presence of nonlinear integral operators makes these models quite difficult to deal with both from a theoretical and numerical point of view. For such reason, alternative models have been formulated, after the relaxation one proposed by Bhatnagar, Gross, and Krook (BGK) for a single gas \cite{BGK54}. The main advantage of the BGK formulation is the replacement of the collision contribution by a linear relaxation term, avoiding drawbacks due to nonlinearities induced by the integral operator. The relaxation term drives the evolution towards a proper Maxwellian attractor.
	
	The original BGK model has been extended to multi-constituent mixtures in \cite{AAP} by considering a unique relaxation operator for each mixture component, describing the whole collisional process in one single BGK kernel. Afterward, other BGK models were worked out to suitably reproduce some distinctive features of inert and reactive mixtures of monatomic and polyatomic gases
	\cite{BGS-PhysRevE,Bisi-Monaco-Soares,BT1,BT2,BT3,Brull-Schneider-CMS2014,garzo1989kinetic,Haack-Hauck-Murillo,Klingenberg-Pirner-Puppo}.
	
	In this paper, we consider the BGK model proposed in \cite{BBGSP} for an inert mixture and extended very recently in \cite{MST24} for a reactive mixture. More precisely, the BGK collision operator in \cite{BBGSP} is given by the sum of binary terms, one for each mechanical interaction between any pair of components. In paper \cite{MST24} the authors consider a mixture of four monatomic gases undergoing elastic collisions and a bimolecular reversible chemical reaction \cite{Aracne}. The collision operator of each component preserves the same structure and splits into a sum of BGK terms, considering the mechanical encounters and the chemical interactions separately.
	
	The usual assumption that the exchange rates for momentum and energy of each single BGK mechanical and chemical operator coincide with the corresponding ones of each Boltzmann integral operator is considered here. Then, we deduce a relationship between the auxiliary parameters of local attractors and the main macroscopic fields. 
	The computation of chemical contributions is performed by following the approach proposed in \cite{Kremer-PandolfiBianchi-Soares-PoF2006}, where the input distribution function is assumed to be a proper perturbation of a local Maxwellian with a deviation depending on the constituent fields.
	
	The purpose of this paper is to investigate the role of mechanical and chemical contributions in the relaxation towards the equilibrium, by examining numerically some test cases in space homogeneous conditions. More precisely, we first study a scenario compatible with the result proved in the $H$-theorem presented in \cite{MST24}, then discuss other scenarios when the evolution starts from an initial configuration very far from equilibrium. 
	
	The paper is organized as follows. After recalling the main properties of the model in Section \ref{sec_bgk}, we present the test cases in Section \ref{sec:numerics}. Some concluding remarks are given in Section \ref{sec:concl}.

	\section{A BGK reactive model}
	\label{sec_bgk}
	
	In this section, we review the BGK model proposed in \cite{MST24} and present its main properties. We consider a reactive mixture of four monatomic gases $G_i$, $i=1,2,3,4$, and each component is endowed with its molecular mass $m_i$ and internal energy $E_i$, $i=1,2,3,4$. Particles of the mixture can interact mechanically through binary elastic collisions and undergo a reactive encounter according to the bimolecular reversible chemical reaction
	\begin{equation}
		G_1+G_2 \leftrightarrows G_3+G_4\,.
		\label{reaction}
	\end{equation}
	The chemical reaction results in a rearrangement of masses, so that  
	\begin{equation}
		m_1+m_2=m_3+m_4 \,
		\label{masses}
	\end{equation}
	and in a redistribution of total (kinetic and chemical) energy. We introduce the reaction heat defined by 
	\begin{equation}
		\Delta E = E_3 + E_4 - E_1 - E_2 \,,
		\label{heat}
	\end{equation} 
	and we assume that $\Delta E > 0$, meaning that the forward reaction is endothermic.
	
	
	\subsection{Kinetic equations}
	\label{sec:eqs}
	
	We introduce the distribution  function $f_i(\mathbf{x},\mathbf{v},t)$ of particles of component $i$ in the phase space, depending on space variable $\mathbf{x}\in\mathbb{R}^3$, 
	microscopic velocity $\mathbf{v}\in\mathbb{R}^3$ 
	and time $t\in\mathbb{R}_+$.
	We also introduce the notation $\mathbf{f}=[f_1,f_2,f_3,f_4]$ to indicate the vector collecting all the distribution functions.
	
	The time-space evolution of functions $f_i(\mathbf{x},\mathbf{v},t)$ is described by the following set of kinetic equations
	\begin{equation}
		\dfrac{\partial f_i}{\partial t} + \mathbf{v}\cdot\nabla_{\mathbf{x}}f_i
		=  \sum_{j=1}^4\widetilde{\mathcal{Q}}_{ij} + \widehat{\mathcal{Q}}_i ,
		\quad i=1,2,3,4\,,
		\label{BGK_eq}
	\end{equation}
	where each term $\widetilde{\mathcal{Q}}_{ij}$ and $\widehat{\mathcal{Q}}_i$ ($i,j=1,2,3,4$),
	is a collision operator of BGK type, prescribing the relaxation towards 
	a Gaussian attractor. More precisely, each operator $\widetilde{\mathcal{Q}}_{ij}$ describes mechanical collisions
	between species $i$ and $j$, while the operator $\widehat{\mathcal{Q}}_i$ accounts for the effects due to the chemical interaction.
	
	The relaxation operators are given by
	\begin{equation}
		\widetilde{\mathcal{Q}}_{ij}=\widetilde{\nu}_{ij}\left(\mathcal{M}_{ij}-f_i\right)\qquad \mbox{and} \qquad
		\widehat{\mathcal{Q}}_{i}=\widehat{\nu}_{ij}^{hk}\left(\mathcal{M}_{i}-f_i\right),
	\end{equation}
	where $\widetilde{\nu}_{ij}$ represents the frequency of mechanical collisions between $i$-th and $j$-th particles, 
	and $\widehat{\nu}_{ij}^{hk}$ the frequency of reactive interactions obeying the chemical law (\ref{reaction}).
	Moreover, 
	\begin{equation}
		\mathcal{M}_{ij} = \widetilde n_{ij} \, \mathbf M_i(\mathbf{v};\widetilde{\mathbf{u}}_{ij},\widetilde{T}_{ij})
		\qquad \mbox{and} \qquad
		\mathcal{M}_{i} = \widehat n_i \, \mathbf M_i(\mathbf{v};\widehat{\mathbf{u}}_{i},\widehat{T}_{i}) ,
		\label{eq:atract}
	\end{equation}
	with
	\begin{equation}
		\mathbf M_i(\mathbf{v};\mathbf{u},T)
		= \left(\dfrac{m_i}{2\pi T}\right)^\frac32\exp\left[-\dfrac{m_i}{2T}\left(\mathbf{v}-\mathbf{u}\right)^2\right] ,
		\label{eq:MaxMx}
	\end{equation}
	are Gaussian attractors depending on some fictitious fields 
	$\widetilde{n}_{ij}$, $\widetilde{\mathbf{u}}_{ij}$, $\widetilde{T}_{ij}$, 
	and $\widehat{n}_i$, $\widehat{\mathbf{u}}_i$, $\widehat{T}_i$.
	
	Such fields have been explicitly determined in \cite{MST24}, assuming that the production rates of mass, momentum, and energy associated with each individual mechanical and reactive interaction are the same when considering both the Boltzmann operator and the corresponding BGK one. For the sake of brevity, we present here the resulting expressions for these parameters
	\begin{itemize}
		\item[--] mechanical auxiliary fields
		\begin{equation}
			\begin{aligned}
				\widetilde{n}_{ij}&=n_i\\
				\widetilde{\mathbf{u}}_{ij}&=(1-a_{ij})\mathbf{u}_i+a_{ij}\mathbf{u}_j\\
				\widetilde{T}_{ij}&=(1-b_{ij})T_i+b_{ij}T_j+\gamma_{ij}|\mathbf{u}_i-\mathbf{u}_j|^2
			\end{aligned}
			\label{fict_mech}
		\end{equation}
		being
		\begin{equation}
			a_{ij}=\dfrac{\widetilde{\lambda}_{ij}}{\widetilde{\nu}_{ij}}\alpha_{ji}n_j\,,\quad b_{ij}=2a_{ij}\alpha_{ij}\,,\quad\gamma_{ij}=\dfrac13m_ia_{ij}(2\alpha_{ji}-a_{ij})\,,
		\end{equation}
		where $\widetilde{\lambda}_{ij}$ is the momentum of the cross-section of Boltzmann formulation, assumed to be constant because of the Maxwell molecule hypothesis \cite{maxwell1866xiii}, and $\alpha_{ij}=m_i/(m_i+m_j)$ denotes the mass ratio;
		
		\bigskip
		
		\item[--] chemical auxiliary fields
		\begin{equation}
			\begin{aligned}
				\widehat{n}_{i}&=n_i+\dfrac{\widehat{P}_{i}^{(0)}}{m_i\widehat{\nu}_{ij}^{hk}}\\
				\widehat{\mathbf{u}}_{i}&=\dfrac{1}{\widehat{n}_i}\left(n_i\mathbf{u}_i+\dfrac{\widehat{P}_{i}^{(1)}}{m_i\widehat{\nu}_{ij}^{hk}}\right)\\
				\widehat{T}_{i}&=\dfrac{1}{\widehat{n}_i}\left\{n_iT_i+\dfrac23\left[\dfrac{\widehat{P}_{i}^{(2)}}{\widehat{\nu}_{ij}^{hk}}-\dfrac12m_i\left(\widehat{n}_i\widehat{\mathbf{u}}_i^2-n_i\mathbf{u}_i^2\right)-E_i\left(\widehat{n}_i-n_i\right)\right]\right\}\,,
			\end{aligned}\label{fict_chem}
		\end{equation}
		where $\widehat{P}_{i}^{(0)}$, $\widehat{P}_{i}^{(1)}$, $\widehat{P}_{i}^{(2)}$ are the Boltzmann production terms for mass, momentum, and energy, respectively, explicitly computed in detail in \cite{MST24}.
	\end{itemize}
	
	\medskip
	
	\noindent
	Therefore, by construction, the BGK model guarantees the correct conservation laws and chemical exchange rates. 
	
	We observe that the positivity of auxiliary mechanical temperatures is guaranteed under the condition $\widetilde{\nu}_{ij}\ge\widetilde{\lambda}_{ij}n_j/2$, see \cite{BBGSP}. 
	Instead, the positivity of chemical number densities and temperatures presents some technical difficulties to be overcome.
	Anyway, assuming an initial condition sufficiently close to the equilibrium, it is possible to prove this by using continuity arguments.
	
	\medskip
	
	Closing this section, we highlight a significant feature of the BGK model, namely its capability to describe separately the effects of chemical reactions and mechanical interactions, allowing for the analysis of different chemical regimes of evolution. Furthermore, since the mechanical collision operator decomposes into several contributions, each one describing the effects of a pair of constituents, the model also details the mechanical effects due to exchanges between components within the collisional dynamics. Therefore, the BGK model considered here is well equipped, by construction, to efficiently describe both the mixture effects and the influence of the chemical reaction.
	
	
	\subsection{Relaxation to equilibrium and entropy estimate}
	\label{sec:trend}
	
	The steady state of a reactive mixture requires both mechanical and chemical equilibrium, and it is assured by distribution functions
	for which both collision operators in equation (\ref{BGK_eq}) simultaneously vanish. 
	This leads to Maxwellian distribution functions sharing the same mean velocity and temperature, that is
	\begin{equation}
		f_i^M(\mathbf{v}) 
		= n_i \, \mathcal{M}(\mathbf{v};\mathbf{u},T)
		= n_i\left(\dfrac{m_i}{2\pi T}\right)^{\!\frac32} \exp\left[-\dfrac{m_i}{2T}\left(\mathbf{v}-\mathbf{u}\right)^2\right] ,
		\label{eq:MaxMx}
	\end{equation}
	with the additional constraint that species number densities fulfill the mass action law
	\begin{equation}
		\left(\dfrac{m_1m_2}{m_3m_4}\right)^{\!\frac32} n_3n_4\exp\left(\dfrac{\Delta E}{T}\right)-n_1n_2=0\, .
		\label{MAL}
	\end{equation}
	In a space homogeneous configuration, and under specific conditions, see \cite{MST24}, the global stability of such an equilibrium state has been proven, employing the usual Lyapunov functional for reactive kinetic systems,
	\begin{equation}
		\mathcal{H}[\mathbf{f}] 
		= \sum_{i=1}^4\int_{\mathbb{R}^3}f_i(\mathbf{v})
		\log \left(\dfrac{f_i(\mathbf{v})}{m_i^3}\right)d\mathbf{v} .
		\label{eq:Hchem}
	\end{equation}
	More precisely, the main result is the following.
	

	\begin{theorem}
		\label{theorem_H}
		Let us assume that auxiliary parameters of chemical     
		attractors given in (\ref{eq:atract})
		satisfy the conditions
		\begin{equation}
			{\widehat{\mathbf{u}}{_i}} = \widehat{\mathbf{u}},
			\qquad
			\widehat{T}_i=\widehat{T}
			\label{equal_cond}
		\end{equation}
		and
		\begin{equation}
			\dfrac{\widehat{n}_1\widehat{n}_2}{\widehat{n}_3\widehat{n}_4}
			=  \left(\dfrac{m_1m_2}{m_3m_4}\right)^{\!\frac32} \exp\left(\dfrac{\Delta E}{\widehat{T}}\right) .
			\label{MAL_chem}
		\end{equation}
		Under space homogeneous conditions, for all measurable distribution functions $f_i\ge 0$, $i=1,2,3,4$, we have that
		
		\medskip
		
		(a) $\dfrac{d\mathcal{H}}{d t}\le 0\,, \quad \text{for all} \;\; t\ge 0$;
		
		\bigskip
		
		(b) $\dfrac{d{\cal H}}{dt}(t) = 0 \qquad \mbox{if and only if} \qquad f_i = f_i^M, \ \ \mbox{for}  \quad i=1,2,3,4$.
		
		\label{th:HH}
	\end{theorem}
	
	
	
	\section{Numerical experiments}
	\label{sec:numerics}
	
	In this section, we conduct some numerical simulations to examine the approach to equilibrium of a reacting mixture, under space homogeneous conditions, for varying initial configurations. Following the framework outlined in \cite{GS2004}, our analysis is performed under the additional assumption of isotropic distribution functions, i.e. 
	$f_i(\mathbf v)=f_i(v)$, that leads to $\mathbf{u}_i=\mathbf{u}=\mathbf{0}$.
	
	The numerical values of parameters and initial conditions are chosen just for illustrative purposes and have to be considered as dimensionless quantities, without any reference to a specific scenario. 
	
	The macroscopic fields, as the moments of the distribution functions, are computed in spherical coordinates and evaluated through a proper trapezoidal quadrature scheme. The resulting system of partial differential equations is discretized in the modulus of microscopic velocity, and the resulting discrete system of ODEs is solved numerically using a Matlab routine based on an adaptive Runge-Kutta method.

	\subsection{Asymptotic trend to equilibrium}
	
	The first scenario considered here aims to support numerically the main result proved in \cite{MST24} and recalled in Theorem \ref{theorem_H}, i.e. the existence of an H-functional, and hence the asymptotic approach	to equilibrium, starting from a close-to-equilibrium configuration. More precisely,
	we suppose that each component of the mixture has its initial distribution function of Maxwellian shape, which writes
	\begin{equation}
		f_i^0(\mathbf{v}) 
		= n_i^0\left(\dfrac{m_i}{2\pi T_i^0}\right)^{\!\frac32} \exp\left[-\dfrac{m_i}{2T^0_i}\left({v}\right)^2\right]\,.
		\label{eq:MaxMx}
	\end{equation}
	
	We choose as masses and internal energies of the four components the following ones
	\begin{equation}\label{masse}
		m_1=9.7,\quad m_2=5.6,\quad m_3=8,\quad m_4=7.3,
	\end{equation}
	\begin{equation}\label{energie} 
		E_1=8\times10^{-4},\quad E_2=6\times10^{-4},\quad E_3=1.3\times10^{-3},\quad E_4=1.1\times10^{-3}.
	\end{equation}
	Then, we fix, as initial densities and temperatures, the following
	\begin{equation}\label{densIn1}
		n_1^0=1	,\quad n_2^0=1.2,\quad n_3^0=1.4,\quad n_4^0=1.3,
	\end{equation}
	\begin{equation}\label{tempIn1}   
		T_1^0=4\times 10^{-2},\quad T_2^0=4.3\times 10^{-2},\quad T_3^0=3.7\times 10^{-2},\quad T_4^0=3.5\times 10^{-2}.
	\end{equation}
	To compute the auxiliary fields for BGK operators defined through relations \eqref{fict_mech}-\eqref{fict_chem}, we set momenta of the cross section and collision frequencies as follows
	\begin{equation}
		\tilde{\lambda}_{ij}=0.001,\quad i,j=1,2,3,4,\quad
		\tilde{\nu}_{ij}=\left(\begin{array}{cccc}
			3 & 4 & 1 & 4 \\
			4 & 3 & 4 & 6 \\
			1 & 4 & 3 & 2 \\
			4 & 6 & 2 & 4
		\end{array}\right),
	\end{equation}
	\begin{equation}
		\hat{\nu}_{ij}^{hk}=1,\quad (i,j),(h,k)\in\big\{(1,2),(3,4),(2,1),(4,3)\big\}, i\neq h,i\neq k.
	\end{equation}
	
	The numerical simulations show the relaxation of distribution functions to Maxwellian equilibria, sharing the same temperature, when the initial configuration is close to an equilibrium one, even if the assumptions \eqref{equal_cond} in Theorem \ref{theorem_H} are not imposed. The usual H-functional for chemically reacting mixtures given in equation (\ref{eq:Hchem}), is monotone also in this case, and the entropy estimate is still guaranteed.
	
\begin{figure}[ht!]
		\centering
		\includegraphics[scale=0.25]{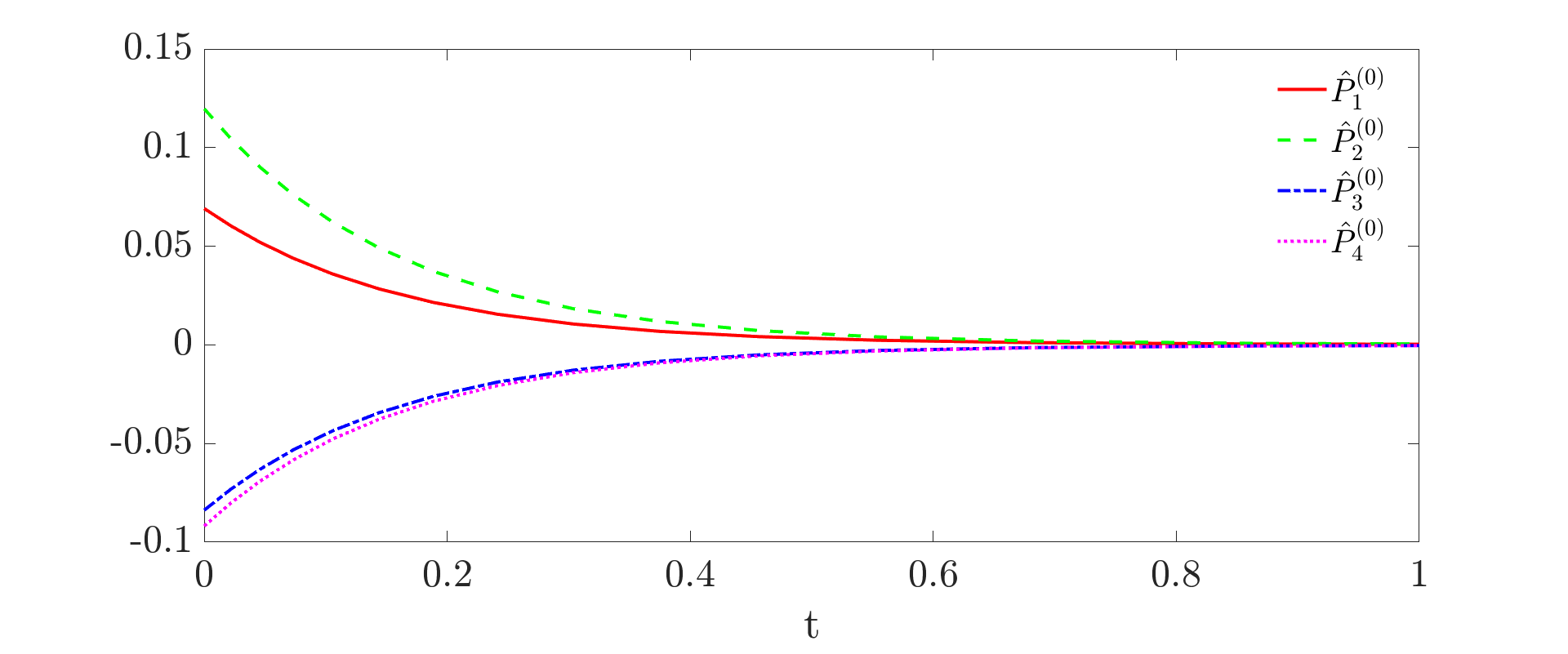}
		\includegraphics[scale=0.25]{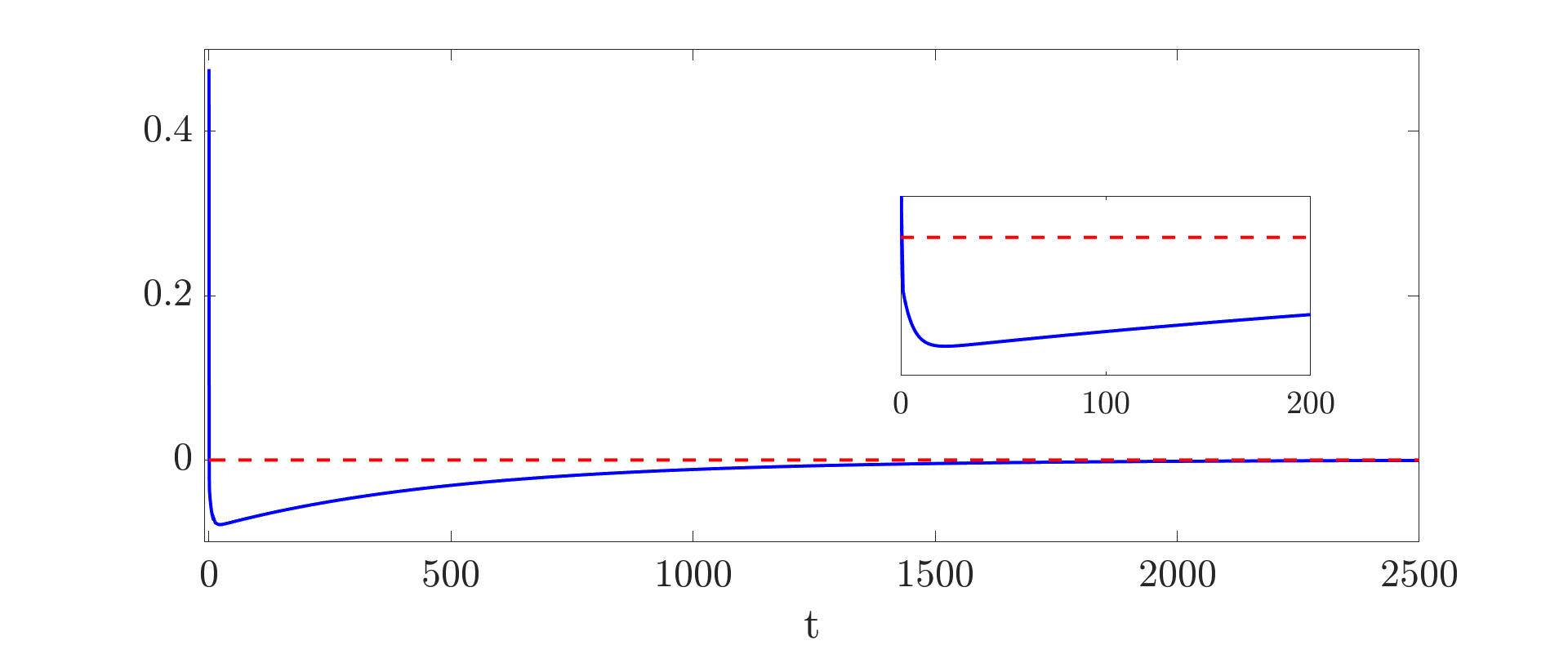}
		\caption{{\it Scenario 1 -- Trend to equilibrium.} Chemical production terms for mass (top) and deviation from chemical equilibrium measured by the left-hand side of the mass action law (\ref{MAL}) (bottom).}
		\label{prod_MAL}
	\end{figure}

	The simulations allow us to investigate in detail the role of chemical contributions. In particular, 
	Figure \ref{prod_MAL} (top) shows that the production terms for mass due to chemical exchanges between components approach monotonically to $0$. This behavior indicates that the mass action law given in (\ref{MAL}), characterizing the equilibrium from a chemical point of view, is asymptotically satisfied. Figure \ref{prod_MAL} (bottom) shows the deviation from the chemical equilibrium measured by the expression on the left-hand side of the mass action law (\ref{MAL}).  This figure shows that this deviation crosses the null value in a very early stage of the evolution, but it continues decreasing since the components have not reached the mechanical equilibrium yet.
	
\begin{figure}[ht!]
		\centering
		\includegraphics[scale=0.25]{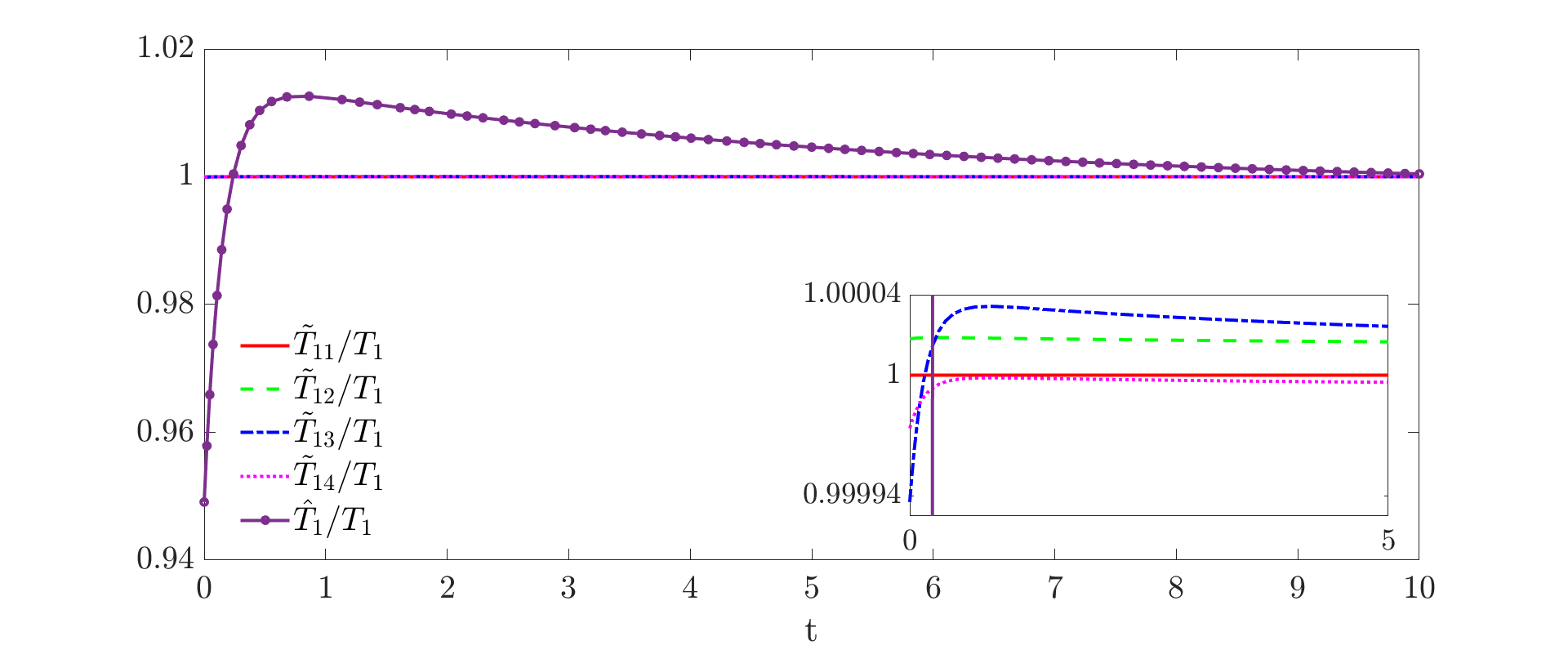}
		\includegraphics[scale=0.25]{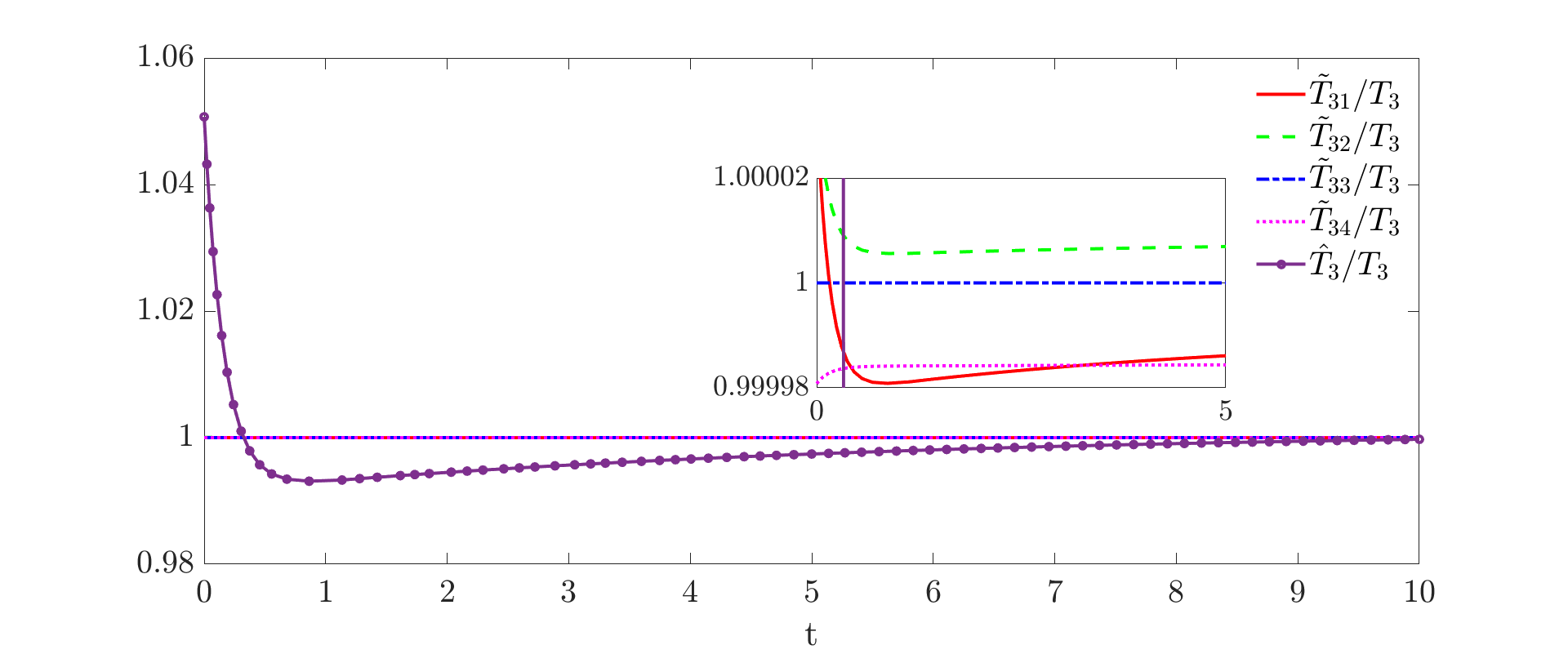}
		\caption{{\it Scenario 1 -- Trend to equilibrium.} Auxiliary temperatures in mechanical and chemical BGK terms, rescaled with respect to species temperature, for the first component (top) and the third component (bottom).}
		\label{aux_temp}
	\end{figure}

	The behavior represented in Figure \ref{prod_MAL} is in agreement with that depicted in Figure \ref{aux_temp}, where it is shown that the relaxation of chemical auxiliary temperatures $\hat{T}_i$ to $T_i$ ($i=1$ top, $i=3$ bottom) occurs later than the convergence of mechanical fictitious ones $\tilde{T}_{ij}$ to $T_i$. Also in this plot, we notice that the profile of $\hat{T}_i/T_i$ crosses the value 1 very early in the transient, but it does not assume this value definitively, since the auxiliary temperatures in mechanical BGK terms still differ from each other.

\begin{figure}[ht!]
		\centering
		\includegraphics[scale=0.25]{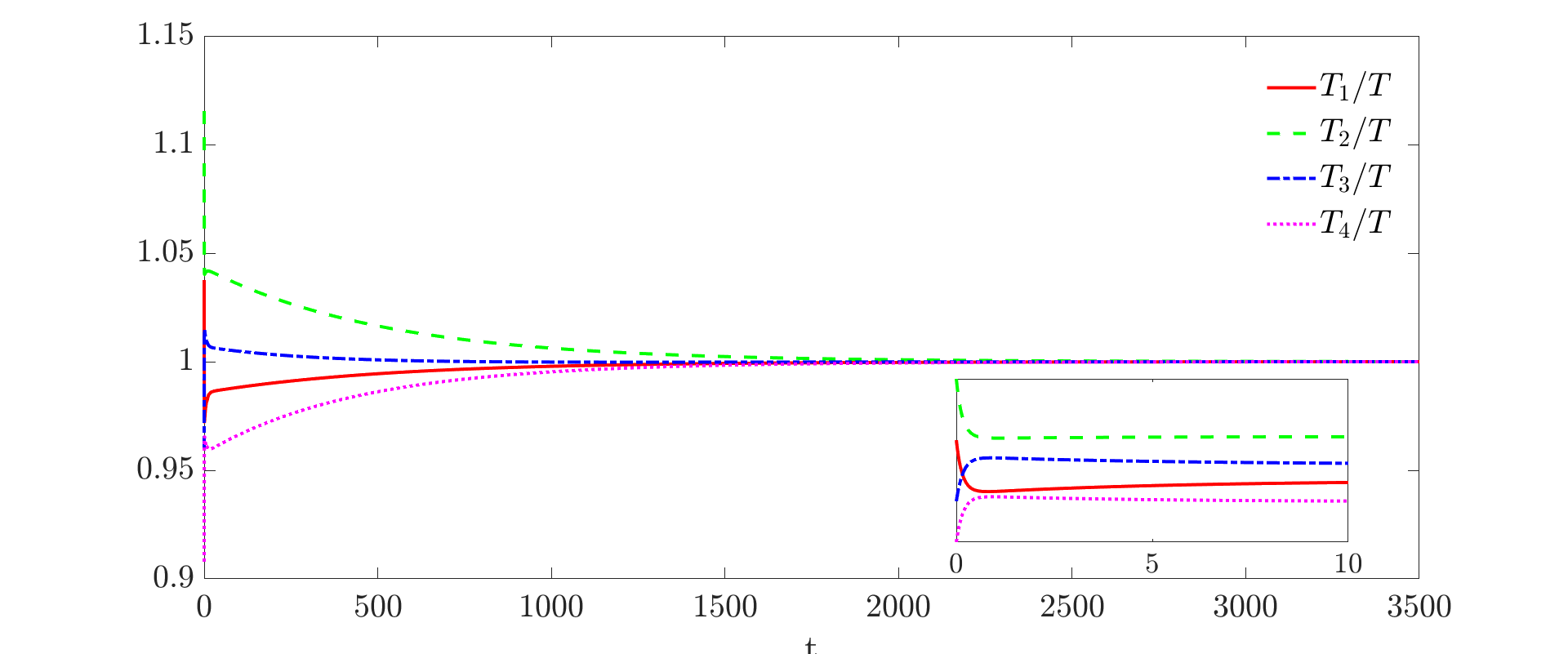}
		\caption{{\it Scenario 1 -- Trend to equilibrium.} Species temperatures scaled with respect to the global temperature.}
		\label{T_vere}
	\end{figure}

	A last comment for this scenario regards the trend to the equilibrium of species temperatures scaled with respect to the mixture temperature, depicted in Figure \ref{T_vere}. Once again, both mechanical and chemical exchanges between components play their own role in the transient, where we can observe non-monotone behaviors of the species profiles before the equalization stage, where	the temperatures relax to equilibrium.	
	
	

	%
	
	
	\subsection{Far from equilibrium case}
	
	As a second scenario, we consider initial distribution functions having a triangular shape (piecewise linear functions), 
	as already proposed in the literature for analogous modelings \cite{GS2004}, 
	supposing that all molecules are initially confined to a compact support. 
	Widths and heights of these linear splines are picked in such a way they provide the following initial densities and temperatures
	\begin{equation}\label{densIn2}
		n_1^0=2.9\times 10^{-2},\quad n_2^0=1.5\times 10^{-1},\quad n_3^0=9.7\times 10^{-1},\quad n_4^0=3.9\times 10^{-1},
	\end{equation}
	\begin{equation}\label{tempIn2}   
		T_1^0=9\times 10^{-2},\quad T_2^0=7.5\times 10^{-2},\quad T_3^0=16.6\times 10^{-2},\quad T_4^0=5\times 10^{-3}.
	\end{equation}
	All the remaining parameters are set as in the previous scenario. We underline that initial distribution functions are significantly far from the expected Maxwellian profile at equilibrium, 
	and species temperatures differ sensitively from each other.
	
\begin{figure}[ht!]
		\centering
		\includegraphics[scale=0.25]{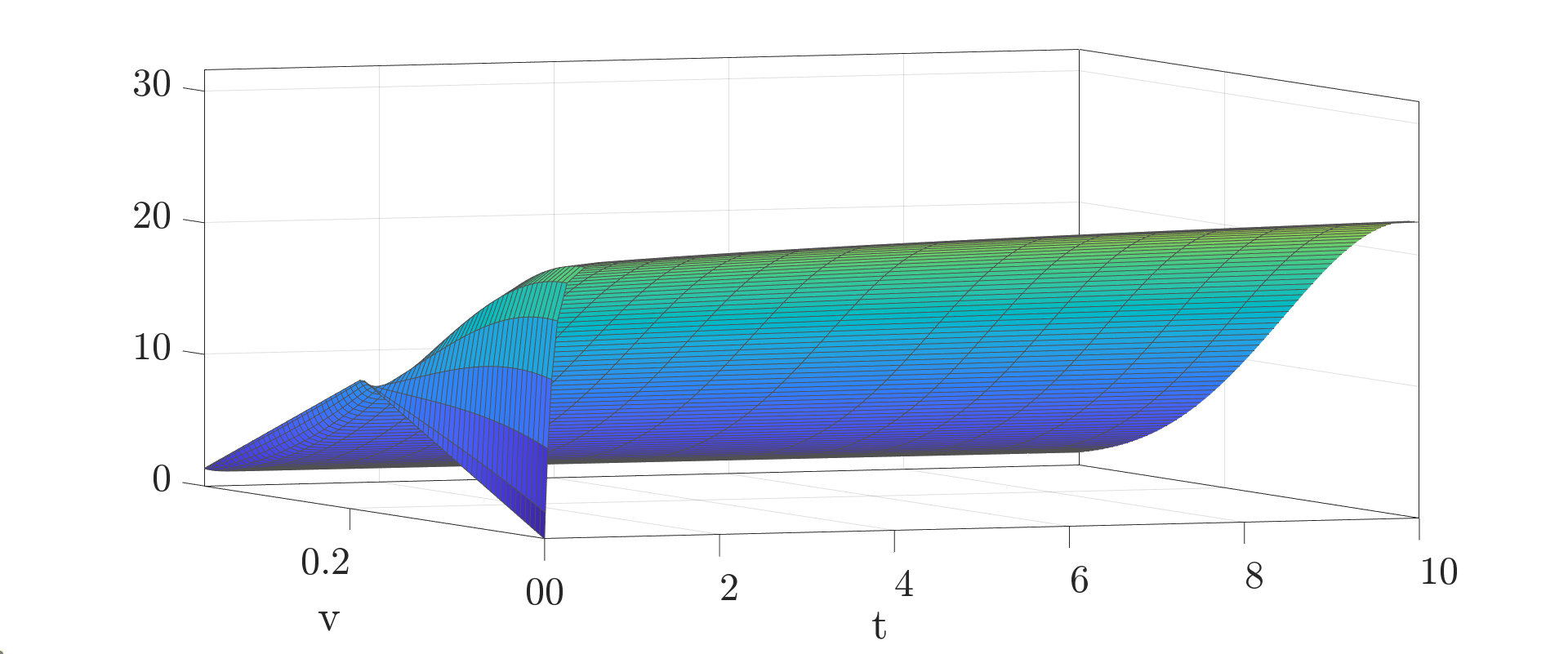}
		\includegraphics[scale=0.25]{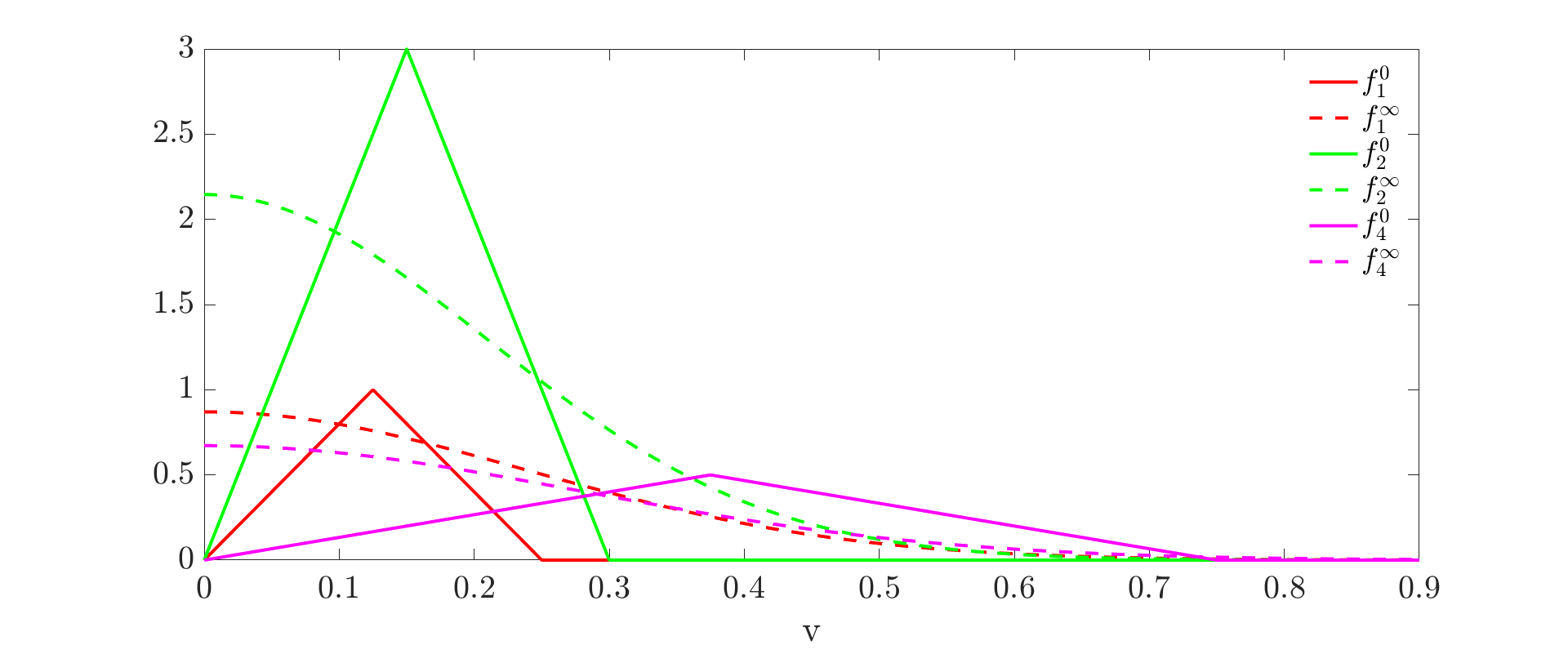}
		\caption{{\it Scenario 2 -- Far from equilibrium.} 
			Distribution function $f_3$ in $(t,v)-$plane (top). Initial distribution $f_i^{0}$
			and asymptotic distribution functions $f_i^{\infty}$ for species 1,2 and 4 (bottom).}
		\label{dist_func}
	\end{figure}

	In Figure \ref{dist_func}, we can see that each distribution function approaches a proper Maxwellian and the exchanges between components have a smoothing effect on the piecewise initial distribution.
	Moreover, we can observe that all the constituents reach a common value of the final temperature, 
	being the variances of the Gaussian profiles all the same.

\begin{figure}[ht!]
		\centering
		\includegraphics[scale=0.25]{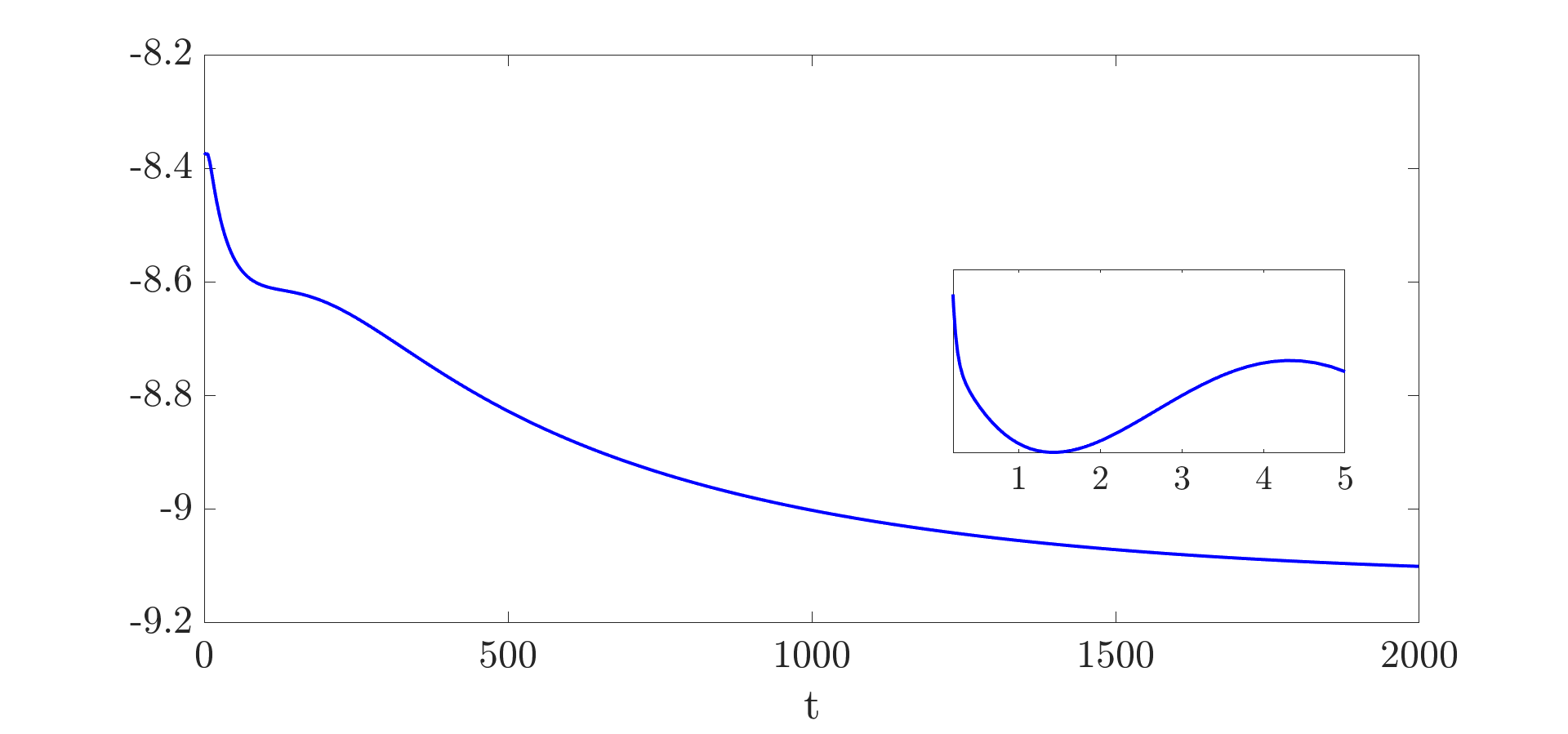}
		\caption{{\it Scenario 2 -- Far from equilibrium.}
			H-functional given in (\ref{eq:Hchem}).}
		\label{H_func}
	\end{figure}

	It is important to underline that the relaxation to equilibrium cannot be proved by means of the usual H-functional for reacting mixtures, since, as shown in Figure \ref{H_func}, the H-functional given in (\ref{eq:Hchem}) decreases only at a later stage, exhibiting a non-monotone behavior and two peaks in the initial transient (see the zoom).
	
\begin{figure}[ht!]
		\centering
		\includegraphics[scale=0.25]{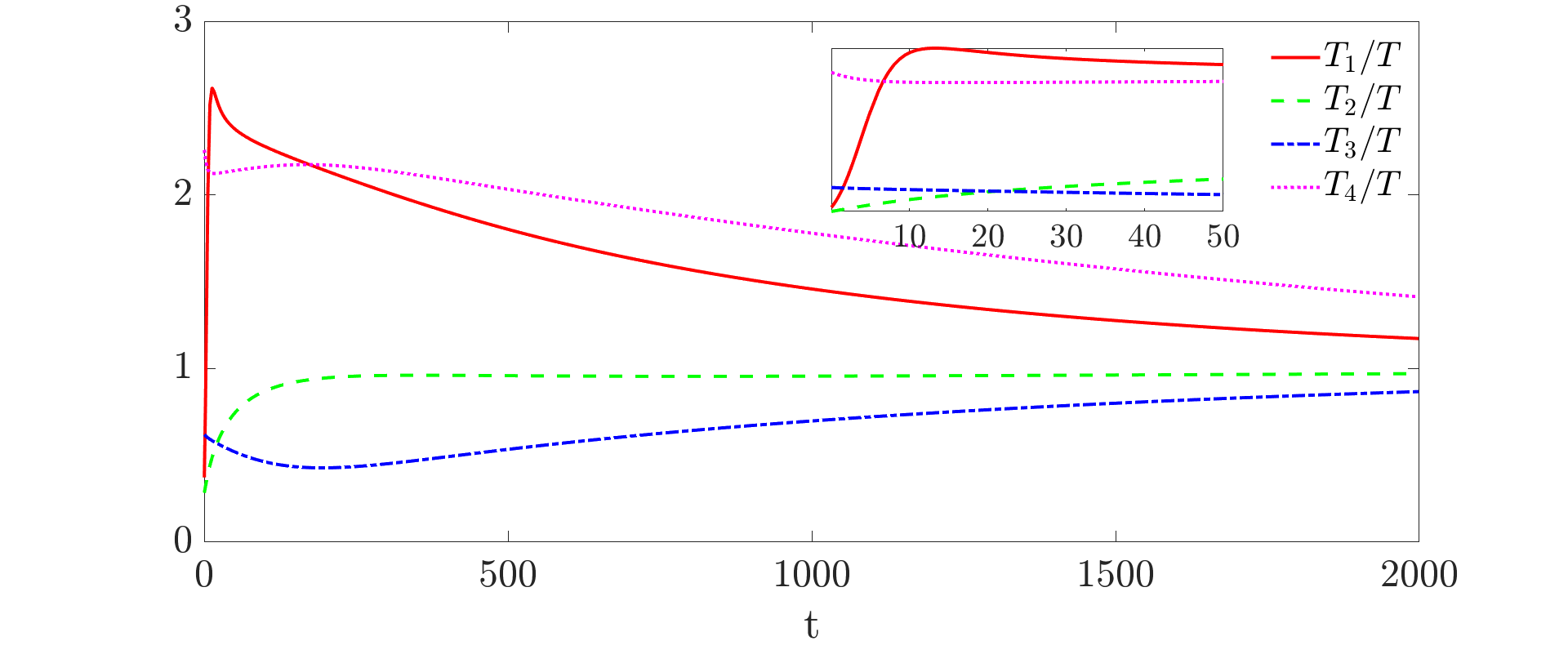}
		\caption{{\it Scenario 2 -- Far from equilibrium.}
			Species temperatures scaled with respect to the global temperature.}
		\label{T_vere_2}
	\end{figure}

	The approaching of species temperatures to equilibrium is represented in Figure \ref{T_vere_2}. As expected, being the initial configuration now far from the equilibrium, the reaching time is longer compared to the one observed in the previous scenario. We notice that the heavier species (red profile) presents a steep slope at the beginning, probably due to the combination of a high value of mass and a low density. More precisely, the high value of $m_1$ pushes $T_1$ towards the mixture temperature $T$, overcomes the threshold, and reaches the global temperature value from above. Also, $T_4$ presents a similar behavior, probably due to the initial value. In fact, deviations between this species temperature and the others are higher in production terms, significantly pushing the species value towards the mixture one.
	
\begin{figure}[ht!]
		\centering
		\includegraphics[scale=0.25]{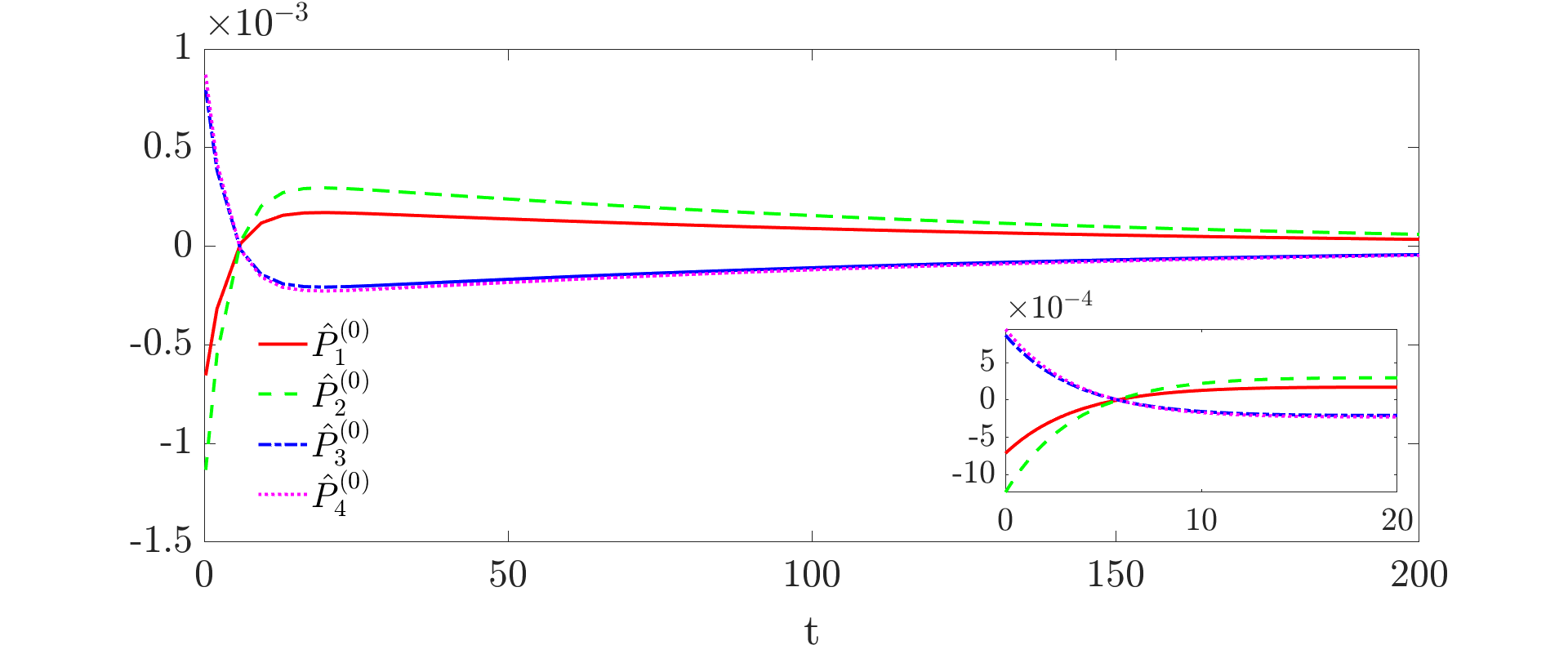}
		\includegraphics[scale=0.25]{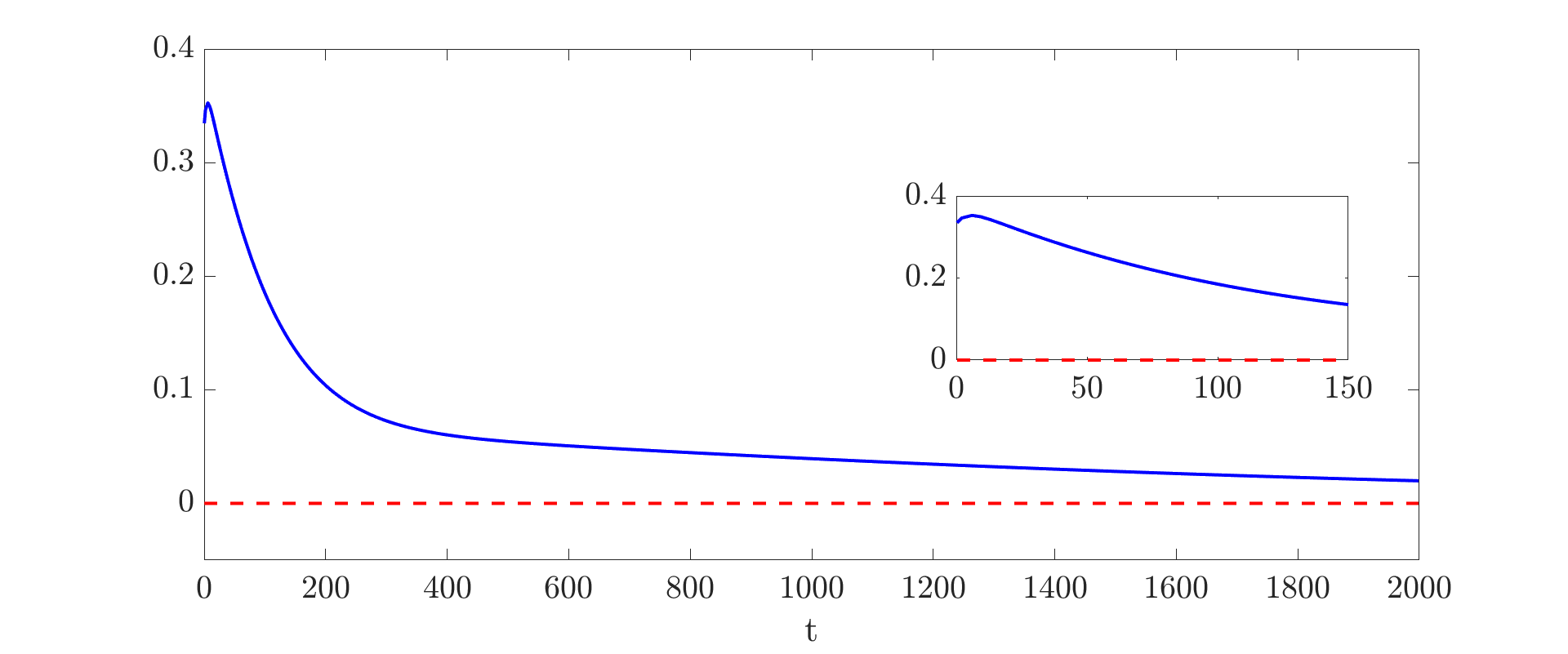}
		\caption{{\it Scenario 2 -- Far from equilibrium.}
			Chemical production terms for mass (top) and left-hand side of (\ref{MAL}) (bottom).}
		\label{prod_MAL_2}
	\end{figure}

	Lastly, a few comments are due concerning the role of chemical contributions in the evolution.
	As in the previous scenario, the chemical exchange terms influence the first phase of the transient. 
	This second scenario results in being richer than the first one. In particular, species production terms, represented in Figure \ref{prod_MAL_2} (top), reach the null value simultaneously but they do not assume this value definitely, since components are still far from the mechanical equilibrium. Moreover, Figure \ref{prod_MAL_2} (bottom) shows that the deviation from equilibrium measured by the	left-hand side of the mass action law (\ref{MAL}) initially exhibits an increasing behavior and, after a peak, it decreases towards 0.
	
	\section{Conclusions}
	\label{sec:concl}
	
	We have considered a recent BGK model \cite{MST24} extending the BGK model proposed \cite{BBGSP} to the reactive framework, with the aim of investigating the asymptotic convergence to equilibrium. The theoretical analysis developed in \cite{MST24} proves an H-theorem prescribing an entropy estimate, for a class of distribution functions, by considering a hypothesis of quasi-equilibrium configuration.
	
	The numerical investigation clearly shows that the usual H-function for reacting mixtures is a Lyapunov functional in a wider class. Moreover, we have shown the relaxation to equilibrium also when the initial distribution functions are far from the equilibrium, even if the H-functional is not globally non-increasing.
	
	We have discussed the contribution of chemical terms in the evolution, by underlying that the exchanges between components are crucial, especially in the transient. We have also observed that the relaxation of chemical fictitious temperatures to species temperature, and hence to the global one of the mixture, occurs later than the equalization of auxiliary mechanical temperatures.
	
	A future work will regard the extension of such analysis to space-dependent problems, like the evaporation-condensation, and the Riemann problem, as well as the comparison with other models for chemically reacting mixtures \cite{BGS-PhysRevE,Brull-Schneider-CMS2014}.
	
	The present paper, coupled with our previous paper \cite{MST24}, could be inspiring to extend a recent hybrid Boltzmann-BGK model, that has been proposed in \cite{Ferrara,BGLM} for an inert mixture of monatomic gases, to a reactive mixture, possibly with a polyatomic structure, and to investigate complex processes of reactive systems in different evolution regimes.  
	
	
	\paragraph{Acknowledgments.}
	$\,$
	G.M. and R.T. thank the Italian National Group of Mathematical Physics (GNFM-INdAM) 
	and University of Parma (Italy).
	The authors thank the support of Cost Action CA18232.
	R.T. and A.J. thank the support of
	Portuguese FCT-CMAT-UM Projects  UIDB/00013/2020 (\url{https://doi.org/10.54499/UIDB/00013/2020}), 
	UIDP/00013/2020 (\url{https://doi.org/10.54499/UIDP/00013/2020}).\\
	G.M. thanks the support of the Italian PRIN 2022 PNRR Research Project
	``Mathematical Modelling for a Sustainable Circular Economy in Ecosystems''
	(P2022PSMT7) funded by the European Union-NextGenerationEU and by MUR-Italian Ministry of Universities and Research.\\
	A.J. thanks for the support of PT-FR Pessoa Project Ref.2021.09255.CBM.\\
	R.T. is a post-doc fellow supported by the National Institute of Advanced Mathematics (INdAM), Italy.
	R.T. thanks the support of
	Bando di Ateneo 2022 per la ricerca co-funded by MUR-Italian Ministry of Universities and 
	Research - D.M. 737/2021-PNR-PNRR-NextGenerationEU, 
	Project ``Collective and self-organised dynamics: kinetic and network approaches''.



\begin{thebibliography}{6}
	
	
	\bibitem{AAP}
	P. Andries, K. Aoki, B. Perthame,
	A consistent BGK-type model for gas mixtures,
	\emph{J. Stat. Phys.}, \textbf{106}, 993--1018, 2002.
	
	
	\bibitem{BGK54}
	P.L. Bhatnagar, E.P. Gross, K. Krook,
	A model for collision processes in gases,
	{\em Phys. Rev.}, \textbf{94}, 511--524, 1954.
	
	
	\bibitem{Ferrara}
	M. Bisi, W. Boscheri, G. Dimarco, M. Groppi, G. Martal\`o,
	A new mixed Boltzmann-BGK model for mixtures solved with an IMEX finite volume scheme on unstructured meshes,
	\emph{Appl. Math. Comput.}, \textbf{433}, 127416, 2022. 
	
	
	\bibitem{BGLM}
	M. Bisi, M. Groppi, E. Lucchin, G. Martal\`o,
	A mixed Boltzmann-BGK model for inert gas mixtures,
	\emph{Kinet. Relat. Models}, \textbf{17}, 674--696, 2024.
	
	
	
	
	\bibitem{Aracne}
	M. Bisi, M. Groppi, G. Spiga,
	Kinetic modelling of bimolecular chemical reactions,  
	in {\em Kinetic Methods for Nonconservative and Reacting Systems}, G. Toscani ed.,
	Quaderni di Matematica, \textbf{16}, 1--143,
	Aracne Editrice, Roma, 2005.
	
	
	
	\bibitem{BGS-PhysRevE}
	M. Bisi, M. Groppi, G. Spiga,
	Kinetic Bhatnagar--Gross--Krook model for fast reactive mixtures and its hydrodynamic limit,
	\emph{Phys. Rev. E}, \textbf{81}, 036327 1--9, 2010.
	
	
	\bibitem{Bisi-Monaco-Soares}
	M. Bisi, R. Monaco, A.J. Soares,
	A BGK model for reactive mixtures of polyatomic gases with continuous internal energy,
	\emph{J. Phys. A - Math. Theor.}, \textbf{51}, 125501 1--29, 2018.
	
	
	\bibitem{BT1}
	M. Bisi, R. Travaglini, A BGK model for mixtures of monoatomic and polyatomic gases with discrete internal energy, 
	{\em Phys. A: Stat. Mech. Appl.}, \textbf{547}, 124441, 2020.
	
	
	\bibitem{BT2}
	M. Bisi, R. Travaglini, 
	A kinetic BGK relaxation model for a reacting mixture of polyatomic gases, 
	in {\em Recent Advances in Kinetic Equations and Applications}, 
	F. Salvarani ed., Springer INdAM Series, 
	\textbf{48}, Springer International Publishing, 2021.
	
	
	\bibitem{BT3}
	M. Bisi, {R. Travaglini}, 
	BGK model for a mixture with two reversible reactions, 
	in {\em INdAM Workshop The Legacy of Carlo Cercignani: from Kinetic Theory to Turbulence Modeling}, 
	Springer Nature Singapore, 59--72, 2021.
	
	
	
	\bibitem{BBGSP}
	A.V. Bobylev, M. Bisi, M. Groppi, G. Spiga, I.F. Potapenko,
	A general consistent BGK model for gas mixtures,
	{\em Kinet. Relat. Models}, \textbf{11}, 1377--1393, 2018.
	
	
	
	
	\bibitem{Brull-Schneider-CMS2014}
	S. Brull, J. Schneider,
	Derivation of a BGK model for reacting gas mixtures,
	\emph{Commun. Math. Sci.}, \textbf{12}, 1199--1223, 2014.
	
	
	
	
	
	
	
	\bibitem{garzo1989kinetic}
	V. Garz{\'o}, A. Santos, J.J. Brey, 
	A kinetic model for a multicomponent gas,
	\emph{Phys. Fluids},
	\textbf{1}(2), {380--383},{1989}.
	
	
	
	\bibitem{GS2004}
	M. Groppi, G. Spiga,
	A Bhatnagar–Gross–Krook-type approach for chemically reacting gas mixtures,
	{\em Phys. Fluids}, \textbf{16}, 4273--4284, 2004.
	
	
	\bibitem{Haack-Hauck-Murillo}
	J.R. Haack, C.D. Hauck, M.S. Murillo,
	A conservative, entropic multispecies BGK model,
	{\em J. Stat. Phys.}, \textbf{168}, 826--856, 2017
	
	
	\bibitem{Klingenberg-Pirner-Puppo}
	C. Klingenberg, M. Pirner, G. Puppo,
	A consistent kinetic model for a two--component mixture with an application to plasma,
	{\em Kinet. Relat. Models}, \textbf{10}, 445--465, 2017.
	
	
	\bibitem{Kremer-PandolfiBianchi-Soares-PoF2006}
	G.M. Kremer, M. Pandolfi Bianchi, A.J. Soares,
	A relaxation kinetic model for transport phenomena in a reactive flow,
	{\em Phys.~Fluids}, \textbf{18}, 037104 1--15, 2006.
	
	
	\bibitem{MST24}
	G. Martal\`o, , A.J. Soares, R. Travaglini,
	A BGK-type model for multi-component gas mixtures undergoing a bimolecular chemical reaction, 2024 ({\em submitted}).
	
	
	\bibitem{maxwell1866xiii}
	J.C. Maxwell,
	XIII. The Bakerian Lecture. -- On the viscosity or internal friction of air and other gases,
	\emph{Philos. Trans. Royal Soc}, \textbf{156}, 249--268, 1866.
	
	
	
	
	
	
	
	\end{thebibliography}
\end{document}